\documentclass[12pt]{article}

\usepackage{amsmath,amssymb,amsfonts,amsbsy}
\usepackage{cite}
\usepackage{graphicx}
\usepackage{wrapfig}

\begin{document}
\addcontentsline{toc}{subsection}{{Orbital momentum effects due to  a
liquid nature of transient state}\\
{\it S.M.Troshin}}

\setcounter{section}{0}
\setcounter{subsection}{0}
\setcounter{equation}{0}
\setcounter{figure}{0}
\setcounter{footnote}{0}
\setcounter{table}{0}

\begin{center}
\textbf{ORBITAL MOMENTUM EFFECTS DUE TO  A LIQUID NATURE OF
TRANSIENT STATE}

\vspace{5mm}

\underline{S.M. Troshin} and N.E. Tyurin

\vspace{5mm}

\begin{small}
   \emph{Institute for High Energy Physics, 142281, Protvino, Moscow Region, Russia}
\end{small}
\end{center}

\vspace{0.0mm} 

\begin{abstract}
It is argued that directed flow $v_1$, the observable introduced
for description of nucleus collisions, can be used for the
detection of the nature of state of the matter in the transient
state of hadron and nuclei collisions.  We consider a possible
origin of the directed flow in hadronic reactions as a result of
rotation of the transient matter and trace analogy with nucleus
collisions. Our proposal it that the presence of directed flow can
serve as a signal that transient matter is in a liquid state.
\end{abstract}

\vspace{7.2mm}
 Important tools in the studies of the nature of the
new form of matter are the anisotropic flows which are the
quantitative characteristics of the collective motion of the
produced hadrons in the nuclear interactions. With their
measurements one can obtain a valuable information on  the early
stages of reactions and observe signals of QGP formation. The
experimental probes  of collective dynamics in $AA$ interactions,
the momentum anisotropies $v_n$  are defined by means of the
Fourier expansion of the transverse momentum spectrum over the
momentum azimuthal angle $\phi$. The angle $\phi$ is the angle of
the detected particle transverse momentum with respect to the
reaction plane spanned by the collision axis $z$ and the impact
parameter vector $\mathbf b$
 directed along the $x$ axis. Thus, the anisotropic flows are  the
 azimuthal correlations with the reaction plane.
In particular, the directed flow is defined as
\begin{equation}\label{dirfl}
v_1(p_\perp)\equiv \langle \cos \phi \rangle_{p_\perp}= \langle
{p_x}/{p_\perp}\rangle = \langle {\hat{\mathbf  b}\cdot {\mathbf
p}_\perp} /{p_\perp}\rangle
\end{equation}
From Eq. (\ref{dirfl}) it is evident that this observable can be
used for studies of multiparticle production dynamics in hadronic
collisions provided that impact parameter $\mathbf  b$ is fixed.

We assume that the origin of the transient state and its dynamics
along with hadron structure can be related to the mechanism of
spontaneous chiral symmetry breaking ($\chi$SB) in QCD,
 which  leads
to the generation of quark masses and appearance of quark
condensates. This mechanism describes transition of the current
into  constituent quarks. The  gluon field is considered to be
responsible for providing quarks with
  masses and its internal structure through the instanton
  mechanism of the spontaneous chiral symmetry breaking.
Massive  constituent quarks appear  as quasiparticles, i.e.
current quarks and the surrounding  clouds  of quark--antiquark
pairs which consist of a mixture of quarks of the different
flavors.  Quark  radii are
  determined by the radii of  the  surrounding clouds.  Quantum
numbers of the constituent quarks are the same as the quantum
numbers of current quarks due to conservation of the corresponding
currents in QCD.

  Collective excitations of the condensate are the Goldstone bosons
and the constituent quarks interact via exchange of the Goldstone
bosons; this interaction is mainly due to pion field. Pions
themselves are the bound states of massive quarks. The interaction
responsible for quark-pion interaction
 can be written in the form \cite{diak}:
 \begin{equation}
{\cal{L}}_I=\bar
Q[i\partial\hspace{-2.5mm}/-M\exp(i\gamma_5\pi^A\lambda^A/F_\pi)]Q,\quad
\pi^A=\pi,K,\eta.
\end{equation}
The interaction is strong, the corresponding coupling constant is
about 4. The  general form of the total effective Lagrangian
(${\cal{L}}_{QCD}\rightarrow {\cal{L}}_{eff}$)
 relevant for
description of the non--perturbative phase of QCD
 includes the three terms \cite{gold} \[
{\cal{L}}_{eff}={\cal{L}}_\chi +{\cal{L}}_I+{\cal{L}}_C.\label{ef}
\] Here ${\cal{L}}_\chi $ is  responsible for the spontaneous
chiral symmetry breaking and turns on first. The picture of a
hadron consisting of constituent quarks embedded
 into quark condensate implies that overlapping and interaction of
peripheral clouds   occur at the first stage of hadron
interaction. The interaction of the condensate clouds assumed to
of the shock-wave type, this condensate clouds interaction
generates the quark-pion transient state. This mechanism is
inspired by the shock-wave production process proposed by
Heisenberg long time ago. At this stage,  part of the effective
lagrangian ${\cal{L}}_C$ is turned off (it is turned on again in
the final stage of the reaction). Nonlinear field couplings
transform then the kinetic energy to internal energy. As a result
the massive virtual quarks appear in the overlapping region and
transient state of matter is generated. This state consist of
$\bar{Q}Q$ pairs and pions strongly interacting with quarks. This
picture of quark-pion interaction
 can be considered as an
origin for percolation mechanism of deconfinement resulting in the
liquid nature of transient matter \cite{jtt}.

Part of hadron energy carried by the outer condensate clouds
being released in the overlap region goes to generation of massive
quarks interacting by pion exchange
 and their number was estimated as follows as follows:
\begin{equation} \tilde{N}(s,b)\,\propto
\,\frac{(1-\langle k_Q\rangle)\sqrt{s}}{m_Q}\;D^{h_1}_c\otimes
D^{h_2}_c \equiv N_0(s)D_C(b), \label{Nsbt}
\end{equation} where $m_Q$ -- constituent quark mass, $\langle k_Q\rangle $ --
average fraction of hadron  energy carried  by  the constituent
valence quarks. Function $D^h_c$ describes condensate distribution
inside the hadron $h$ and $b$ is an impact parameter of the
colliding hadrons. Thus, $\tilde{N}(s,b)$ quarks appear in
addition to $N=n_{h_1}+n_{h_2}$ valence quarks.

The generation time of the transient state $\Delta t_{tsg}$ in
this picture obeys to the inequality
\[
\Delta t_{tsg}\ll \Delta t_{int},
\]
where $\Delta t_{int}$ is the total interaction time. The newly
generated massive virtual quarks play a role of scatterers for the
valence quarks in elastic scattering; those quarks are transient
ones in this process: they are transformed back into the
condensates of the final hadrons.

Under construction of the model  for elastic scattering it was
assumed that the valence quarks located in the central part of a
hadron are scattered in a quasi-independent way off the transient
state
 with interaction radius of valence quark  determined
by  its inverse mass:
\begin{equation}\label{rq}
R_Q=\kappa/m_Q.
\end{equation}
The elastic scattering $S$-matrix in the impact parameter
representation is written in the model in the form of linear
fractional transform:
\begin{equation}
S(s,b)=\frac{1+iU(s,b)}{1-iU(s,b)}, \label{um}
\end{equation}
where $U(s,b)$ is the generalized reaction matrix, which is
considered to be an input dynamical quantity similar to an input
Born amplitude  and related to the elastic scattering scattering
amplitude through an algebraic equation which enables one to
restore unitarity. The function $U(s,b)$  is chosen in the model
as a product of the averaged quark amplitudes
\begin{equation} U(s,b) = \prod^{N}_{Q=1} \langle f_Q(s,b)\rangle
\end{equation} in accordance  with assumed quasi-independent
nature  of the valence quark scattering. The essential point here
is the rise with energy of the number of the scatterers  like
$\sqrt{s}$. The $b$--dependence of the function $\langle f_Q
\rangle$  has a simple form $\langle
f_Q(b)\rangle\propto\exp(-m_Qb/\xi )$.

These notions can be extended to particle production with account
of the geometry  of the overlap region and  properties of the
liquid transient state. Valence constituent quarks would excite a
part of the cloud of the virtual massive quarks  and those quark
droplets will subsequently hadronize  and form the multiparticle
final state.   This mechanism can be relevant for the region of
moderate transverse momenta while the region of high transverse
momenta should be described by the excitation of the constituent
quarks themselves and application of  the perturbative QCD to the
parton structure of the constituent quark. The model allow to
describe elastic scattering and the main features of multiparticle
production. In particular, it leads to asymptotical dependencies
\begin{equation}\label{tota}
  \sigma_{tot,el}\sim \ln^2 s,\;\;
 \sigma_{inel}\sim \ln s, \;\; \bar{n}\sim s^\delta.
\end{equation}
The  geometrical picture of hadron collision at non-zero impact
parameters described above implies that the generated massive
virtual  quarks in overlap region will obtain large initial
orbital angular momentum at high energies. The total orbital
angular momentum  can be estimated as follows
\begin{equation}\label{l}
 L(s,b) \simeq \alpha b \frac{\sqrt{s}}{2}D_C(b).
\end{equation}
The parameter $\alpha$ is related to the fraction of the initial
energy carried by the condensate clouds which goes to rotation of
the quark system and the overlap region, which is described by the
function $D_C(b)$, has an ellipsoidal form. It should be noted
that $L\to 0$ at $b\to\infty$ and $L=0$ at $b=0$. At this point we
would like to stress again on the liquid nature of transient
state. Namely due to strong interaction between quarks in the
transient state, it can be described as  a quark-pion liquid.
Therefore, the orbital angular momentum $L$ should be realized  as
a coherent rotation of the quark-pion liquid  as a whole  in the
$xz$-plane (due to mentioned strong correlations between particles
presented in the liquid). It should be noted that for the given
value of the orbital angular momentum $L$
 kinetic energy has a minimal value if all parts of liquid rotates with the same angular velocity.
  We  assume therefore
that the different parts of the quark-pion liquid in the overlap
region indeed have the same angular velocity $\omega$. In this
model spin of the polarized hadrons has its origin in the rotation
of matter hadrons consist of. In contrast, we assume rotation of
the matter during intermediate, transient state of hadronic
interaction. Collective rotation of the strongly interacting
system of the massive constituent quarks and pions is  the main
point of the proposed
  mechanism of the directed flow generation in hadronic and nuclei collisions.
We  concentrate on the effects of this rotation and consider
directed flow for the constituent quarks supposing that  directed
flow for  hadrons is close to the directed flow for the
constituent quarks at least qualitatively. The assumed particle
production mechanism at moderate transverse momenta is an
excitation of  a part of the rotating transient state of  massive
constituent quarks (interacting by pion exchanges) by the one of
the valence constituent quarks
 with  subsequent hadronization
of the quark-pion liquid droplets. Due to the fact that the
transient matter is strongly interacting, the excited parts should
be located closely  to the periphery  of the rotating transient
state otherwise absorption
 would not allow
to quarks and pions to leave the region (quenching). The mechanism
is sensitive
 to the particular
rotation direction and the directed flow should have  opposite
signs for the particles in the fragmentation regions of the
projectile and target respectively. It is evident that the effect
of  rotation (shift in  $p_x$ value ) is most significant in the
peripheral part of the rotating quark-pion liquid and is to be
weaker in the less peripheral regions (rotation with the same
angular velocity $\omega$), i.e. the directed flow $v_1$ (averaged
over all transverse momenta)  should be proportional to the
inverse depth $\Delta l$ where the excitation of the rotating
quark-pion liquid takes place. The geometrical picture of hadron
collision has an apparent analogy with collisions of nuclei and it
should be noted that the appearance of large orbital angular
momentum should be expected in the overlap region in the
non-central nuclei collisions.
 And then due to strongly interacting
nature of the transient matter we assume that this orbital angular
momentum realized as a coherent rotation of liquid. Thus, it seems
that  underlying  dynamics  could be   similar to the dynamics of
the directed flow in hadron collisions.

We can go further and extend the production mechanism from hadron
to nucleus case also. This extension cannot be straightforward.
First, there will be no unitarity corrections for the anisotropic
flows and instead of valence constituent quarks, as a projectile
we should consider nucleons, which would excite rotating quark
liquid. Of course, those differences will result in significantly
higher values of directed flow. But, the general trends in its
dependence on the collision energy, rapidity of the detected
particle and transverse momentum, should be the same. In
particular, the directed flow in nuclei collisions as well as in
hadron reactions will depend on  the rapidity difference
$y-y_{beam}$ and not on the incident energy. The mechanism
 therefore can provide a qualitative explanation of the incident-energy scaling
 of $v_1$ observed at RHIC \cite{jmpe}.

\end{document}